\documentclass[aps,prd,showpacs,amssymb]{revtex4}

\usepackage{graphicx}
\usepackage{bm}

\begin{document}
\title{Renormalized scalar propagator around a dispiration}
\author{V. A. De Lorenci}
 \email{delorenci@unifei.edu.br}
\author{E. S. Moreira Jr.}
 \email{moreira@unifei.edu.br}
\affiliation{Instituto de Ci\^encias,
Universidade Federal de Itajub\'a,
Av.\ BPS 1303 Pinheirinho, 37500-903 Itajub\'a, MG, Brazil}

\date{December, 2002}

\begin{abstract}
The renormalized Feynman propagator for a scalar field
in the background of a cosmic dispiration (a disclination plus a screw
dislocation) is derived,
opening a window to investigate vacuum polarization effects
around a cosmic string with dislocation, as well as in the bulk of
an elastic solid carrying a dispiration.
The use of the propagator is illustrated by computing vacuum fluctuations. 
In particular it is shown that the  
dispiration polarizes the vacuum giving rise to an
energy momentum tensor which, as seen from a local inertial  frame,
presents non vanishing off-diagonal components.
Such a new effect resembles that where an induced vacuum current
arises around a needle solenoid carrying a magnetic flux 
(the Aharonov-Bohm effect), and may have physical consequences.  
Connections with a closely related background,
namely the spacetime of a spinning cosmic string,
are briefly addressed.

\end{abstract}
\pacs{04.62.+v, 61.72.Lk, 11.27.+d}
\maketitle

\section{Introduction}
\hspace{0.5cm}
Computations on quantum fields in locally flat backgrounds with linear
defects not only are an interesting matter of principle but may also
be useful in physical contexts as different as cosmology
(the gravitational background of a cosmic string \cite{vil94}) and
condensed matter physics (the effective geometry associated
with linear defects in solids \cite{kro81,kat92}).
The nontrivial global geometry of the locally flat background
around a linear defect
induces vacuum polarization
\cite{dav82,ful89,wal94}.
Casimir-like effects near cosmic strings and in the bulk
of a solid carrying a linear defect have been investigated throughout
the years in a variety of situations
(see e.g. \cite{hel86,lin87,fro87,mor00,bor01}).
The literature addresses mainly conical defects
appearing  in connection with the geometrical background
of an ordinary cosmic string \cite{vil94}, and of a disclination in a
continuous elastic solid \cite{kro81,kat92}.

This work begins an investigation of quantum field theory effects
in a spacetime with a cosmic dispiration.
Classical fields were recently considered in Ref. \cite{mor02},
and investigations on quantum theory in related backgrounds
have been carried out in Refs. \cite{bez87,pon98,lor01,lor02}.
In Ref. \cite{gal93},
it has been conjectured that the geometry associated with a cosmic dispiration
may correspond to the gravitational field of certain chiral cosmic strings
\cite{bek92}.

The geometry, that arises in the context of general relativity
\cite{gal93} as well as in the Einstein-Cartan theory
\cite{tod94,let95,pun97} is presented in the next Section.
In Sec. III, the proper time expression (see e.g. \cite{dew65}) of the
Feynman propagator for a scalar field
$\phi(x)$ in the four-dimensional spacetime of a
cosmic dispiration is obtained. Additional manipulations
yield  a propagator which is renormalized with respect to
the Minkowski  vacuum. This renormalized propagator is then used in Sec. IV
to evaluate the vacuum fluctuations 
$\langle\phi ^{2}\rangle$  and $\left< T^{\mu}{}_{\nu} \right>$
in the limit of small masses.
Section IV also contains a detailed analysis of the
behavior of  $\langle\phi ^{2}\rangle$
where disclination and screw dislocation effects are confronted.
Final remarks are outlined in Sec. V, including
a short discussion on global hyperbolicity in the context of a
related background.

Throughout the text
$c=\hbar=1$.

\section{The geometry}

The geometry of the spacetime of a cosmic dispiration
(a cosmic string with dislocation \cite{gal93})
is characterized by the Minkowski line element written in cylindrical
coordinates \cite{gal93,tod94},
\begin{equation}
ds^{2}=dt^{2}-dr^2 -r^2 d\varphi^2 - dZ^2,
\label{mmet}
\end{equation}
and by the identification
\begin{equation}
(t,r,\varphi,Z) \sim (t,r,\varphi+2\pi\alpha,Z+2\pi\kappa),
\label{id}
\end{equation}
where $\alpha$ and $\kappa$ are
parameters corresponding to
a disclination and a screw dislocation, respectively.
By defining new space coordinates
$\theta := \varphi / \alpha$ and $z := Z - (\kappa/\alpha)\varphi$,
Eq. (\ref{mmet}) becomes
$ds^2=dt^2 - dr^2 - \alpha^2 r^2 d\theta^2 - (dz + \kappa d\theta)^2$,
and the usual identification
$(t,r,\theta,z) \sim (t,r,\theta +2\pi,z)$ must be observed.
The case for which $\alpha=1$ and $\kappa=0$ clearly corresponds to the
Minkowski spacetime.

In the context of general relativity Eq. (\ref{id}) encodes the fact that
Eq. (\ref{mmet}) hides a curvature singularity on the
symmetry axis. (In the Einstein-Cartan theory there exists
also a torsion singularity on the symmetry axis, when $\kappa\neq 0$.)

\section{The renormalized propagator}
The Feynman propagator for a scalar field with
mass $m$ in the
background described above is a solution of \cite{dav82}
\begin{equation}
\left(\Box_{x}+m^{2}\right)G_{{\cal F}}(x,x')=
-\frac{1}{r}\delta\left(x-x'\right),
\label{fe0}
\end{equation}
where
$\Box _{x}$ is just the d'Alembertian in Minkowski
spacetime written in cylindrical coordinates.
By taking into account Eq. (\ref{id}), the non-trivial
geometry manifests itself only through
\begin{equation}
G_{{\cal F}}(\varphi+2\pi\alpha, Z+2\pi\kappa)=
G_{{\cal F}}(\varphi,Z),
\label{bc1}
\end{equation}
where the other coordinates were omitted.

The eigenfunctions of the operator
$\Box_{x}+m^{2}$, which are regular at $r=0$ 
and satisfy Eq. (\ref{bc1}), are given by
\begin{equation}
\psi_{\omega,\mu,n,\nu}(x)=
\frac{1}{(2\pi)^{3/2}\alpha ^{1/2}}
J_{|n-\nu\kappa|/\alpha}(\mu r)
e^{i[\nu Z- \omega t+(n-\nu\kappa)\varphi/\alpha]},
\label{eig1}
\end{equation}
where $n$ is an integer, $\omega$ and $\nu$ are real numbers,
$\mu$ is a positive real number
and $J_{\sigma}$ denotes a Bessel function of the first kind.
The corresponding eigenvalues are
$E_{\omega ,\mu ,\nu}=
\mu^{2}+\nu^{2}-\omega ^{2}+m^{2}$, and a convenient
(but rather unusual) normalization has been used \cite{mor95}.

Using the Bessel functions completeness relation \cite{arf85}
$$\int_{0}^{\infty}dk\ kJ_{\sigma}(kr)J_{\sigma}(kr')=
\frac{1}{r}\delta(r-r'),$$
and the Fourier expansion (Poisson's formula)
\begin{equation}
\sum_{n=-\infty}^{\infty}\delta(\theta +2\pi n)=
\frac{1}{2\pi}\sum_{n=-\infty}^{\infty}e^{in\theta},
\label{pformula}
\end{equation}
direct application of
$\Box_{x} + m^{2}$ shows that \cite{mor95}
\begin{widetext}
\begin{eqnarray}
G_{{\cal F}}(x,x')=
-i\int_{0}^{\infty}dT\
\sum_{n=-\infty}^{\infty}
\int_{-\infty}^{\infty}d\omega\
\int_{-\infty}^{\infty}d{\nu}
\int_{0}^{\infty}d\mu\ \mu
e^{-iTE_{\omega ,\mu ,\nu}}
\psi_{\omega,\mu,n,\nu}(x)
\psi^{\ast}_{\omega,\mu,n,\nu}(x'),
\label{pro1}
\end{eqnarray}
where the integration over $T$ is regularized by subtracting
an infinitesimal imaginary term from
$E_{\omega ,\mu ,\nu}$.
Evaluating the integrations over $\omega$
and $\mu$ \cite{grad80}, Eq. (\ref{pro1}) yields
\begin{eqnarray}
G_{{\cal F}}(x,x')&=&
-\frac{(i\pi)^{1/2}}{16\pi^{3}\alpha}
\int_{0}^{\infty}\frac{dT}{T^{3/2}}
e^{i\{[r^{2}+r'^{2}-(t-t')^{2}]/4T - m^{2}T\}}
\nonumber\\
&&\times\int_{-\infty}^{\infty}d\nu\
e^{-iT\nu ^{2}+i[(Z-Z')-\kappa(\varphi-\varphi ')/\alpha]\nu}
\sum_{n=-\infty}^{\infty}I_{|n-\nu\kappa|/\alpha}
\left(r r'/2iT\right)
e^{in(\varphi-\varphi')/\alpha},
\label{66}
\end{eqnarray}
where
$I_{\sigma}$ denotes a modified Bessel function of the first kind.

In order to tackle renormalization, a convenient integral representation
for $I_{\sigma}$ \cite{grad80} is used to obtain the equality,
\begin{eqnarray}
\sum_{n=-\infty}^{\infty} I_{|n+\beta|/\alpha}(y)
e^{in\delta/\alpha}
=
\alpha e^{y\cos\delta-i\beta\delta/\alpha}
-\frac{1}{\pi} \int_{0}^{\infty}
\!d\tau\;   e^{-y\cosh \tau} \sum_{n=-\infty}^{\infty}
\sin (|n+\beta|\pi/\alpha)\
e^{-(|n+\beta|\tau-in\delta)/\alpha},
\label{67}
\end{eqnarray}
\end{widetext}
which holds for $\alpha>1/2$
(smaller values can be considered by taking into 
account terms that were omitted),
and is crucial to implementing renormalization and to
performing the
integration over $\nu$ in Eq. (\ref{66}).
(It should be pointed out that the integral representation
for $I_{\sigma}$ mentioned above has 
previously been used in related contexts \cite{shi92,pon98}.)

Using Eq. (\ref{67}) in Eq. (\ref{66}), it results
\begin{equation}
G_{{\cal F}}(x,x')=G_{{\cal F}}^{0}(x,x')+G^{(\alpha,\kappa)}(x,x'),
\label{feynman}
\end{equation}
where $G_{{\cal F}}^{0}(x,x')$ is
the Feynman propagator in the Minkowski spacetime
and
\begin{widetext}
\begin{eqnarray}
&&G^{(\alpha,\kappa)}(x,x'):=
\frac{(i\pi)^{1/2}}{16\pi^{4}\alpha}\int_{0}^{\infty}
\!\!\!\frac{dT}{T^{3/2}}
\ e^{i\{[r^{2}+r'^{2}-(t-t')^{2}]/4T - m^{2}T\}}
\nonumber
\\
&&
\times\!\!
\int_{-\infty}^{\infty}\!\!\! d\nu\
e^{-iT\nu ^{2}+i[(Z-Z')-\kappa(\varphi-\varphi ')/\alpha]\nu}
\!\!\sum_{n=-\infty}^{\infty}\!\!
e^{-in(\varphi-\varphi')/\alpha}
\sin (|n+\nu\kappa|\pi/\alpha)
\int_{0}^{\infty}\!\!\!d\tau\
e^{-(rr'/2iT)\cosh \tau -|n+\nu\kappa|\tau/\alpha}.
\label{69}
\end{eqnarray}
\end{widetext}

Observing the sine factor in Eq. (\ref{69}), for
$\alpha=1$ and $\kappa=0$ Eq. (\ref{69}) vanishes,
leaving in Eq. (\ref{feynman}) only the Minkowski contribution,
as should be.
In order to built up observables with $G_{{\cal F}}(x,x')$ (cf.
the next Section)
its ultraviolet divergences must be eliminated, which can be done
simply by  dropping $G_{{\cal F}}^{0}(x,x')$ in Eq. (\ref{feynman})
(as the background geometry of a cosmic dispiration is locally flat
[cf. Eq. (\ref{mmet})], all the ultraviolet divergences are encapsulated in
the Minkowski contribution \cite{wal94}).

A more workable expression
for the renormalized  propagator
of a massless scalar field
$D^{(\alpha,\kappa)}(x,x')$ can be obtained
by inserting in Eq. ({\ref{69})
$\delta[\lambda-(n+\nu\kappa)/\alpha]$
(obviously accompanied by integration over $\lambda$).
Recalling that $f(x)\delta(x-y)=f(y)\delta(x-y)$, one
uses Eq. (\ref{pformula})
before evaluating the Gaussian integration over $\nu$.
Finally, the integrations over $\lambda$ and $T$
are separately evaluated \cite{grad80}, resulting in
\begin{widetext}
\begin{eqnarray}
D^{(\alpha,\kappa)}(x,x')=
\frac{i}{2\pi^{2}}\!\sum_{n=-\infty}^{\infty}
\!\int^{\infty}_0  \!\!\! d\tau
\frac{\left[\tau^2+\pi^2-(2\pi\alpha n-\Delta\varphi)^2\right]
\left[r^2+r'^2+2rr'\cosh \tau+(\Delta Z-2\pi n\kappa)^2-(\Delta t)^2\right]^{-1}}
{\left\{[\pi(2\alpha n+ 1)-\Delta\varphi]^2+\tau^2\right\}
{\left\{[\pi(2\alpha n- 1)-\Delta\varphi]^2+\tau^2\right\}}},
\label{moises}
\end{eqnarray}
\end{widetext}
where
$\Delta t:=t-t'$, and likewise for $\varphi$ and $Z$.

As $\kappa\rightarrow 0$,
the summation in
Eq. (\ref{moises}) can be evaluated by considering the power series expansion
of $\psi(x)$ (the logarithmic derivative of the gamma function)
and its properties, yielding as the leading contribution
a familiar integral representation \cite{emi94}
for the renormalized scalar propagator
in an ordinary conical background
($\kappa=0$).
On the other hand, as $\kappa\rightarrow\infty$
, the expression for the leading behavior
is obtained  from Eq. (\ref{moises}) by ignoring the summation and
by setting $n=0$, resulting an integral representation which does not
depend on either the parameters characterizing the cosmic dispiration.

\section{Application}
In the following Eq. (\ref{moises}) will be used to calculate vacuum
averages.

\subsection{$\langle\phi ^{2}(x)\rangle$}

Formally (see e.g. \cite{dav82}), the vacuum fluctuation
$\langle\phi ^{2}(x)\rangle$ can be obtained
by setting $x'=x$ in Eq. (\ref{moises}),
and multiplying the resulting expression by $i$,
\begin{widetext}
\begin{eqnarray}
&&\langle\phi^2(r)\rangle
= -\frac{1}{8\pi^2 r^2}\int^{\infty}_{0}d\tau
\frac{1}{(\pi^2+\tau^2)\cosh^2(\tau/2)}
\nonumber
\\
&&-\frac{1}{4\pi^2 r^2}\int^{\infty}_{0}d\tau\sum_{n=1}^{\infty}
\frac{\tau^2-\pi^2(4\alpha^2 n^2-1)}{\left[\pi^2(2\alpha n+ 1)^2
+ \tau^2\right] \left[\pi^2(2\alpha n - 1)^2 +\tau^2\right]
\left[\cosh^2(\tau/2) + (n\pi \kappa/r)^2\right]},
\label{phi2}
\end{eqnarray}
\end{widetext}
which does not depend on the direction of the screw dislocation,
i.e., on the sign of $\kappa$.
The considerations at the end of the previous section
lead to the fact that as $\kappa/r\rightarrow 0$,
Eq. (\ref{phi2}) yields as the leading contribution
\begin{equation}
\langle\phi ^{2}(r)\rangle=
\frac{1}{48\pi^{2}r^{2}}
\left(\alpha^{-2}-1\right),
\label{dphi2}
\end{equation}
which is the  known \cite{smi90}
behavior around
an ordinary cosmic string.
As $\kappa/r\rightarrow \infty$, the leading contribution
is now given by
\begin{equation}
\langle\phi ^{2}(r)\rangle=
-\frac{1}{48\pi^{2}r^{2}},
\label{sphi2}
\end{equation}
where the integral in the first term of the right hand side
of Eq. (\ref{phi2}) was evaluated numerically.
[It is rather curious that Eq. (\ref{sphi2})
follows from Eq. (\ref{dphi2}) by setting 
$\alpha\rightarrow\infty$.]

Apart from theses asymptotic behaviors, the dependence of
$\langle\phi ^{2}(r)\rangle$ on $r$ is nontrivially hidden in
Eq. (\ref{phi2}), requiring numerical analysis.
The plots (where units are omitted) show how
$\langle\phi ^{2}(r)\rangle$
varies with $r$ for various combinations of values of $\alpha$
and $\kappa$.

%
%
\begin{figure}[thp]
\leavevmode
\centering
\includegraphics[scale=0.4]{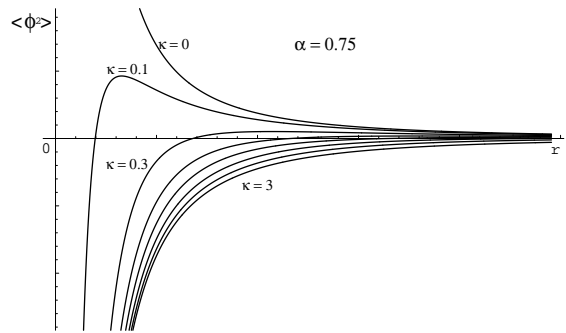}
\caption{The intermediate plots, from the bottom,
correspond to $\kappa=1.5,1,0.7,0.5$, respectively.
\label{conjunto6}
}
\end{figure}
%
%
\begin{figure}[thp]
\leavevmode
\centering
\includegraphics[scale=0.4]{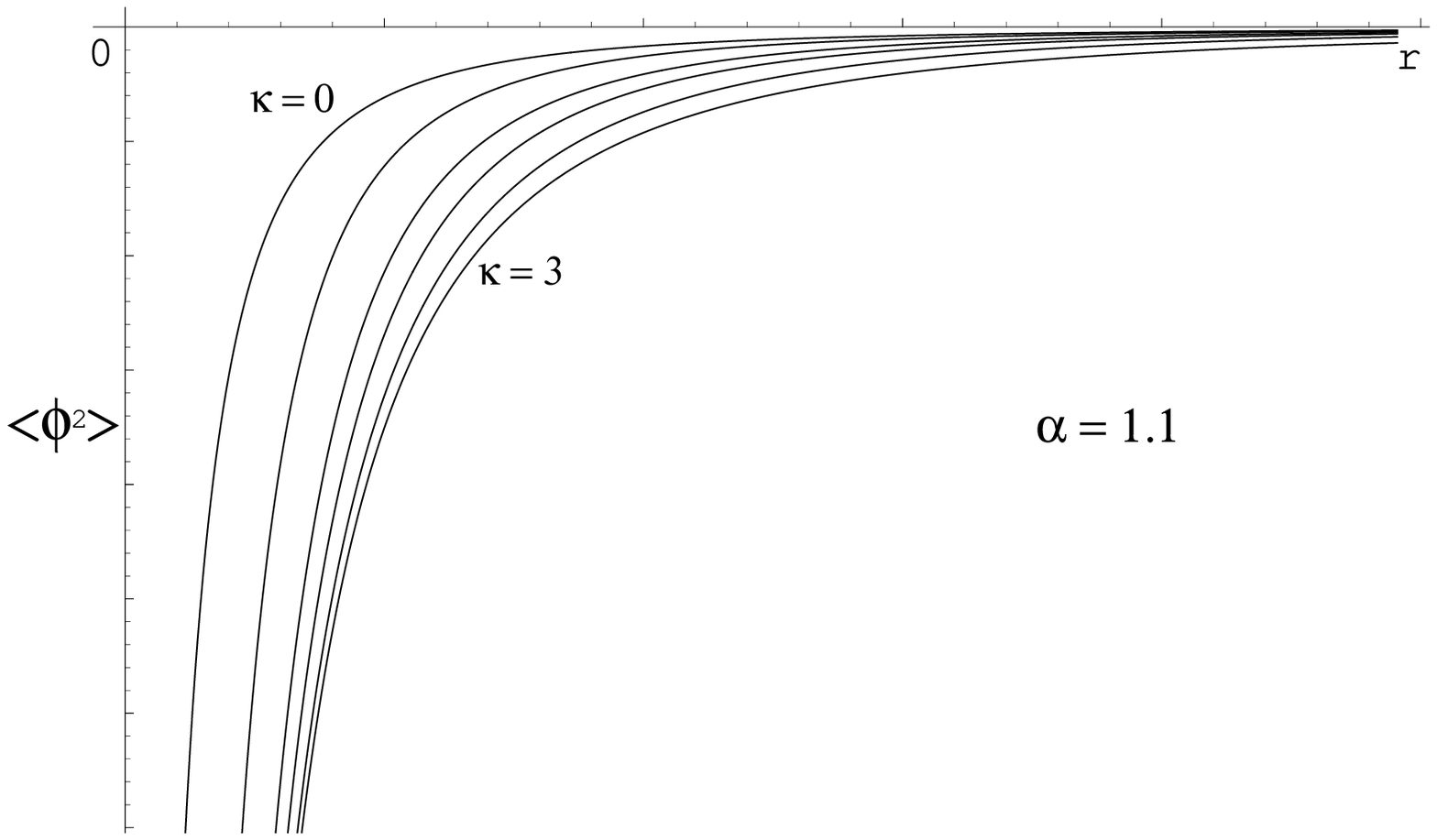}
\caption{The intermediate plots, from the bottom,
correspond to $\kappa=1,0.5,0.3,0.1$, respectively.
\label{conjunto6b}
}
\end{figure}
%
%
\begin{figure}[thp]
\leavevmode
\centering
\includegraphics[scale=0.4]{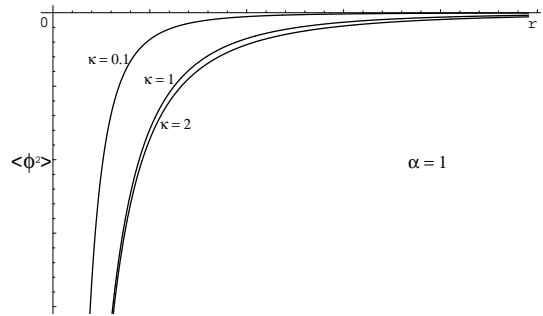}
\caption{Plots corresponding to screw dislocation effects.
\label{conjunto5}
}
\end{figure}
%
%
\begin{figure}[thp]
\leavevmode
\centering
\includegraphics[scale=0.4]{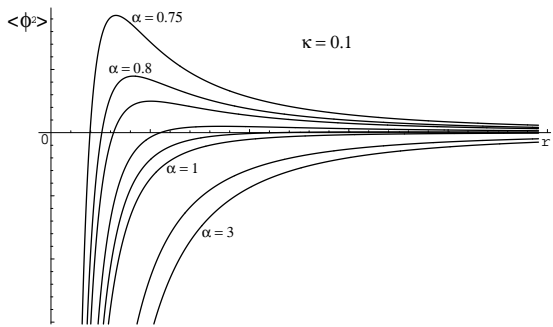}
\caption{The intermediate plots, from the bottom,
correspond to $\alpha=1.5,0.95,0.9,0.83$, respectively.
\label{conjunto3}
}
\end{figure}
%

When $\kappa\neq 0$, Eq. (\ref{dphi2})  shows that
for very large values of $r$ disclination effects
are dominant
(see Figs. \ref{conjunto6} and \ref{conjunto6b}),
whereas
for very small values of $r$,
according to Eq. (\ref{sphi2}),
screw dislocation effects rule
in a way typical of vacuum fluctuations
near boundaries \cite{ful89} (near the dispiration
$\langle\phi ^{2}(r)\rangle$
is essentially independent of the cosmic dispiration attributes).

For arbitrary values of $r$,
numerical and analytical examination
shows that when $\alpha\geq 1$,
$\langle\phi ^{2}(r)\rangle\leq0$
(see Figs. \ref{conjunto6b} and \ref{conjunto5}).
When $\alpha<1$
the screw dislocation and disclination effects compete,
causing $\langle\phi ^{2}(r)\rangle$
to have a maximum which,
by simple dimensional considerations,
is found to be proportional
to $1/\kappa^{2}$, and whose corresponding 
$r=r_{M}$
is proportional to $|\kappa|$ 
(see Figs. \ref{conjunto6} and \ref{conjunto3}).
A similar analysis reveals that, when $\alpha<1$,
$\langle\phi ^{2}(r)\rangle$ 
vanishes at $r=r_{0}$, which is obviously also
proportional to $|\kappa|$.

One notices that, for any $\alpha<1$,
$r_{0}$ and $r_{M}$ move toward $r=0$
as $|\kappa|$ decreases
(see Fig. \ref{conjunto6}).
For a given $\kappa\neq 0$, as $\alpha$
decreases from unity,  
$r_{0}$ and $r_{M}$  also move toward $r=0$
(see Fig. \ref{conjunto3}).

It should be appreciated that when $\alpha<1$ the
$\kappa=0$ and $\kappa\neq 0$ behaviors differ radically from each other
as $r\rightarrow 0$; namely, for $\kappa=0$ 
$\langle\phi ^{2}(r)\rangle$ diverges positively, whereas for
$\kappa\neq 0$   $\langle\phi ^{2}(r)\rangle$ diverges negatively
[see Eq. (\ref{dphi2}), Eq. (\ref{sphi2}) and Fig. \ref{conjunto6}].

It should be pointed out that,
when $\kappa\neq 0$, Eq. (\ref{dphi2})
is the leading contribution as $r\rightarrow \infty$ only when
$\alpha\neq 1$. When $\alpha=1$, Eq. (\ref{dphi2}) vanishes
and the sub-leading contribution, due to the screw dislocation,
takes over. Dimensional considerations may suggest that
such a contribution is proportional to $\kappa^2/r^{4}$.
Nevertheless, an  analysis seems to show that
$\langle\phi ^{2}(r)\rangle$
falls differently
as $r\rightarrow \infty$ (see Figs. \ref{conjunto5} and \ref{conjunto3}).
[In fact, Eq. (\ref{phi2}) is not very handy to determine sub-leading
contributions, and an alternative expression can be more useful for this
purpose.]

\subsection{$\left< T^{\mu}{}_{\nu}(x)\right>$}

As is well known (see e.g. Ref. \cite{dav82})
the vacuum expectation value of the energy momentum tensor can
formally be obtained by applying the differential operator 
\begin{equation}
{\cal D}^\mu{}_\nu(x,x'):=
(1-2\xi)\nabla^\mu\nabla_{\nu'} - 2\xi\nabla^\mu\nabla_{\nu} 
+ (2\xi - 1/2)\delta^\mu{}_\nu
\nabla^\lambda\nabla_{\lambda'} 
\label{doperator}
\end{equation}
to
the renormalized scalar propagator,
\begin{equation}
\left<T^\mu{}_\nu\right> = i \lim_{x'\rightarrow x}
{\cal D}^\mu{}_\nu(x,x')\
D^{(\alpha,\kappa)}(x,x').
\label{prescription}
\end{equation}
By using Eqs. (\ref{moises}) and (\ref{prescription}),
a detailed analytical and numerical study of 
$\left<T^\mu{}_\nu\right>$ (along the lines of that presented above
for $\langle\phi ^{2}\rangle$) is possible,
but rather lengthy to be included here. Only the behavior of 
$\left<T^\mu{}_\nu\right>$ far away from the dispiration 
will be presented below.

As mentioned previously, when $\kappa\rightarrow 0$ the dominant contribution
in Eq. (\ref{moises}) is the renormalized propagator around a disclination.
It follows that  as $\kappa/r\rightarrow 0$,  Eq. (\ref{prescription}) yields
for the diagonal components essentially the expressions long 
known in the literature for the vacuum fluctuations around 
an ordinary cosmic string ($\kappa=0$) \cite{hel86,lin87,fro87}.
Regarding the remaining components, the prescription in Eq.
(\ref{prescription}) kills off the dominant contribution in Eq.
(\ref{moises}), with the result that the subleading contribution yields two 
nonvanishing off-diagonal components,  
\begin{equation}
\left<T^{\varphi}{}_{Z}\right> = \frac{i}{r^2}\lim_{x'\rightarrow x}
\partial_{\varphi}\partial_{Z}D^{(\alpha,\kappa)}(x,x')
= \frac{\kappa}{r^6} B(\alpha)
\label{t23}
\end{equation}
and 
\begin{equation}
\left<T^{Z}{}_{\varphi}\right> 
= \frac{\kappa}{r^4} B(\alpha),
\label{t32}
\end{equation}
where
\begin{equation} 
B(\alpha) := \frac{1}{32\pi^3\alpha^2}\int_0^\infty\! d\tau
\frac{\alpha\sin(\pi/\alpha)\,
[\, \cos(\pi/\alpha) -\cosh(\tau)+
\tau\sinh(\tau)\, ] - 
\pi\, [\, \cos(\pi/\alpha)\cosh(\tau)-1\, ]}
{[\,\cosh(\tau)-\cos(\pi/\alpha)\,]^2\cosh^4(\alpha\tau/2)}.
\label{I}
\end{equation}
It is worth remarking  that, unlike the diagonal components,
$\left<T^{\varphi}{}_{Z}\right>$
and  
$\left<T^{Z}{}_{\varphi}\right>$ 
do not depend on the coupling parameter $\xi$.

The plot of $B(\alpha)$ against the disclination parameter $\alpha$ is
shown in Fig. \ref{fig}. When $\alpha=1$, the
integration in Eq. (\ref{I}) can be analytically evaluated \cite{grad80},
resulting in $B=1/60\pi^{2}$ which
corresponds approximately to the value of $\alpha$ suggested by the physics
of formation of ordinary cosmic strings \cite{vil94}.   
%
\begin{figure}[thp]
\leavevmode
\centering
\hspace*{-0cm}
\includegraphics[scale=0.6]{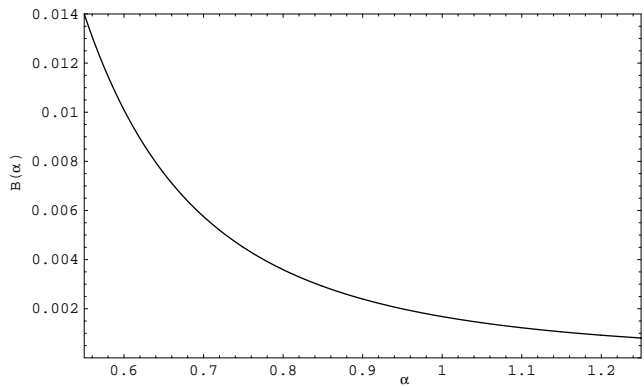}
\caption{ Plot of $B(\alpha)$ versus $\alpha$.
\label{fig}}
\end{figure}
%

It is instructive to display both disclination and screw dislocation
effects in the same array. When $\xi=1/6$ (conformal coupling), for 
example, $\left< T^{\mu}{}_{\nu} \right>$  with respect to the 
local inertial frame [see Eq. (\ref{mmet})] can
be cast into the form  
\begin{equation}
\left< T^{\mu}{}_{\nu} \right> =\frac{1}{r^4}
\left(
     \begin{array}{cccc}
      -A &  0 & 0                    & 0 \\
       0 & -A & 0                    & 0 \\
       0 &  0 & 3A                   & \kappa B/r^2 \\
       0 &  0 & \kappa B             & -A
     \end{array}
\right)
\label{tmunumatrix},
\end{equation}
where $A(\alpha):=(\alpha^{-4}-1)/1440\pi^2$, and which holds 
far away from the defect 
(and for $\alpha\neq 1$, when $\kappa\neq 0$).
[When $\kappa\neq 0$, by setting $\alpha=1$ in Eq. (\ref{tmunumatrix}), 
$A$ vanishes and subleading contributions depending on $\kappa$ take over.]

\section{Final remarks}

This work presented a renormalized expression for
the Feynman propagator of a scalar field 
$\phi(x)$ around a cosmic
dispiration. The propagator
was then used to calculate
$\langle\phi ^{2}(x)\rangle$ and the 
asymptotic behavior of  
$\left< T^{\mu}{}_{\nu}(x) \right>$
as preliminary
exercises to tackle more elaborate vacuum fluctuations.
The expectation is that the results may be useful
in cosmology and condensed matter physics. Few remarks are in order.

It was argued in the previous section that,
as far as $\langle\phi ^{2}(r)\rangle$ is concerned,
disclination effects are dominant over screw dislocation effects
when
$r\rightarrow\infty$, and the other way around  when
$r\rightarrow 0$.
This result should not be extended to all
vacuum fluctuations without caution, since some of them are
obtained from the renormalized propagator
by applying prescriptions
which may eliminate the dominant contribution in Eq.
(\ref{moises}) [see e.g. Eqs. (\ref{t23}) and (\ref{t32})].

The eigenfunctions in Eq. (\ref{eig1}) have the form
$R(r)\chi(\varphi)\exp\{i(\nu Z-\omega t)\}$, 
where
$\chi(\varphi +2\pi\alpha)= \exp\{-i2\pi\nu\kappa\}\chi(\varphi)$. 
This boundary condition is typical of the Aharonov-Bohm setup
where $\nu\kappa$ is identified with the flux parameter $e\Phi/2\pi$.
By carrying over to the four-dimensional context lessons from 
gravity in three dimensions \cite{ger89a,ger90}, it follows that
the charge $e$ and the magnetic flux $\Phi$ should be identified with
the longitudinal linear momentum $\nu$ and $2\pi\kappa$, respectively
\cite{gal93}. (In fact the roots of this analogy lie in the gauge theory
aspects of gravity.) According to this picture,  one interprets the
polarization effect displayed in Eq. (\ref{t32}) in the following manner.
A dispiration (more precisely, a
screw dislocation) polarizes the vacuum of a scalar field, inducing a flux 
of longitudinal linear momentum around the defect. 
Such a flux depends on the direction of the screw dislocation 
(i.e., on the sign of $\kappa$) in the same way that 
vacuum currents around a needle solenoid depend on the direction of the
magnetic flux \cite{ser85}.

Symmetry considerations show that the geometrical
background of a cosmic dispiration is closely related
to that of a spinning cosmic string \cite{gal93},
which is also locally flat [see Eq. (\ref{mmet})] with
a time helical structure characterized by
the identification
$(t,r,\varphi,Z) \sim (t+2\pi S,r,\varphi+2\pi\alpha,Z)$
[instead of the space helical structure in Eq. (\ref{id})].
Such a helical time structure poses serious difficulties when
implementing quantization, since the corresponding spacetime
is not globally hyperbolic \cite{ful89}.
In fact, by insisting on
implementing quantization with the usual procedures, one
ends up with observables presenting pathological behavior
\cite{ger89a,lor01,lor02}. [$\langle\phi ^{2}(r)\rangle$ around a
spinning cosmic string can be obtained from Eq. (\ref{phi2})
by taking $\kappa\rightarrow iS$, rendering the vacuum fluctuation
divergent \cite{lor01}.]

The summation in Eq. (\ref{phi2}) could, in fact, be evaluated.
The resulting integral representation for 
$\langle\phi^{2}(r)\rangle$, however,
does not offer either analytical or computational advantage.
It is given by Eq. (9) of Ref. \cite{lor01}, after replacing
$S$ by $i\kappa$. 
[Incidentally,  it can be noticed that the factor $\sin(x/\alpha)$ in 
Eq. (9) of Ref. \cite{lor01} is a misprint which should be 
read as $\sinh(x/\alpha)$.]

\begin{acknowledgments}
The authors are grateful to Renato Klippert and Ricardo Medina for clarifying
discussions.  This work was partially supported by the Brazilian research
agencies CNPq and FAPEMIG.

\end{acknowledgments}


\end{document}